\pdfoutput=1
\RequirePackage{ifpdf}
\ifpdf 
\documentclass[pdftex]{sigma}
\else
\documentclass{sigma}
\fi

\let\a=\alpha \let\b=\beta  \let\g=\gamma  \let\d=\delta
\let\eps=\epsilon \let\om=\omega \let\m=\mu \let\n=\nu \let\r=\rho \let\s=\sigma

\newcommand{\f}{\frac}

\newcommand{\na}{\nabla}

\newcommand{\tr}{{\rm tr}}
\newcommand{\w}{\wedge}

\newcommand{\R}{\mathbb R}

\newcommand{\scr}{\rm\scriptscriptstyle}

\begin{document}

\allowdisplaybreaks

\renewcommand{\thefootnote}{$\star$}

\renewcommand{\PaperNumber}{032}

\FirstPageHeading

\ShortArticleName{On the Relations between Gravity and BF Theories}

\ArticleName{On the Relations between Gravity and BF Theories\footnote{This
paper is a contribution to the Special Issue ``Loop Quantum Gravity and Cosmology''. The full collection is available at \href{http://www.emis.de/journals/SIGMA/LQGC.html}{http://www.emis.de/journals/SIGMA/LQGC.html}}}

\Author{Laurent FREIDEL~$^\dag$ and Simone SPEZIALE~$^\ddag$}

\AuthorNameForHeading{L.~Freidel and S.~Speziale}

\Address{$^\dag$~Perimeter Institute, 31 Caroline St N, Waterloo ON,  N2L 2Y5, Canada}
\EmailD{\href{mailto:lfreidel@perimeterinstitute.ca}{lfreidel@perimeterinstitute.ca}}

\Address{$^\ddag$~Centre de Physique Th\'eorique, CNRS-UMR 7332, Luminy Case 907, 13288 Marseille, France}
\EmailD{\href{mailto:simone.speziale@cpt.univ-mrs.fr}{simone.speziale@cpt.univ-mrs.fr}}

\ArticleDates{Received January 23, 2012, in f\/inal form May 18, 2012; Published online May 26, 2012}

\Abstract{We review, in the light of recent developments, the existing relations between gravity and topological BF theories at the classical level. We include the Plebanski action in both self-dual and non-chiral formulations, their generalizations, and the MacDowell--Mansouri action.}


\Classification{83C45}

\renewcommand{\thefootnote}{\arabic{footnote}}
\setcounter{footnote}{0}

\section{Introduction}

It is an intriguing fact that general relativity can be given a polynomial action principle which relates it to topological BF theory. The fundamental variable is a connection, and the metric is a derived quantity.
For BF theory alone,  a Lagrange multiplier, $B$, enforces the curvature of the connection to be constant. This type of theories require no metric to be formulated, and posses no local degrees of freedom. The relation with general relativity is established through the extraction of metric degrees of freedom from the fundamental f\/ields. There are mainly two ways of achieving this that have appeared in the literature, based on original work by Plebanski~\cite{Plebanski} and by MacDowell and Mansouri \cite{MacDowell}. The two ways dif\/fer in the choice of gauge group, and in the way the metric is recovered~-- from either the $B$ f\/ield or the connection.
In this review, we will go through the basic aspects of these two approaches, as well as some recent developments.
We restrict our attention to four spacetime dimensions.
For the connection between BF theory and gravity in three dimensions, see \cite{ BlauTFT,CattaneoBF, Witten3d}, and \cite{FreidelExtraD} in dimensions higher than four.

\section{BF theory}

Let us begin with some general aspects of BF theory \cite{BlauTFT}. We consider a principal bundle over the spacetime manifold $M$, with local group a Lie group $G$ and connection $\om$, with curvature~$F(\om)$, and the following action,
\begin{gather}\label{SBF}
S(B,\om) = \int {\rm tr}\, B \w F - \f{\lambda}2 \, {\rm tr} \, B\w B,
\end{gather}
where $B$ is a 2-form in the adjoint representation of $G$ and tr denotes the scalar product in the algebra. $\lambda$ is a dimensionless constant.
Notice that the action is well-def\/ined without any reference to a metric structure on~$M$.
The f\/ield equations are
\begin{gather*}
F(\om)=\lambda B, \qquad d_\om B =0.
\end{gather*}
The action is invariant under local gauge transformations,
\begin{gather*}
\d^G_\a B = [B,\a], \qquad \d^G_\a \om =d_\om\a,
\end{gather*}
and under the following shift symmetry,
\begin{gather}\label{st}
\d^S_\eta B = d_\om \eta, \qquad \d^S_\eta \om =\lambda \eta.
\end{gather}
The total symmetry group is thus the semi-direct product of these two groups, and it is the same for all values of $\Lambda$.
Overall, there are $N$ gauge parameters in the algebra-valued scalars $\a^i$, and $4N$ in the algebra-valued 1-forms $\eta_\mu^i$. The system has so much symmetry that all solutions are locally gauge equivalent, and local degrees of freedom are absent. This is most directly established in the canonical formalism.

The phase space of \eqref{SBF} is spanned by the $6N$ variables $(\om_a^i, \Pi^a_i:=\f12\eps^{abc}B_{ibc})$,
where $a=1,2,3$ are space indices. The action takes the form
\begin{gather*}
S = \int \Pi^a_i \dot{\om}^i_a + \om_0^i {\cal G}_i + B^i_{0a} {\cal C}^a_i,
\end{gather*}
and the  (fully constrained) dynamics is given by the $N$ Gauss constraints ${\cal G}_i$ and $3N$ ``Hamiltonian'' constraints ${\cal C}^a_i$,
\begin{gather}\label{BFcon}
{\cal G}_i=D_a \Pi^a_i = 0,
\qquad {\cal C}^a_{i}= \f12 \eps^{abc} F_{ibc}(\om) -\lambda\Pi^a_i=0.
\end{gather}
As a consequence of the Bianchi identity, $D_a {\cal C}^a_i\equiv 0$, thus there are only $3N$ independent constraints.
The algebra of constraints is
\begin{gather*}
\{C(N), C(N')\} = 0, \qquad \{G(\a), G(\a')\} = G([\a,\a']), \qquad \{C(N), G(\a)\} = C([N,\a]),
\end{gather*}
where $C(N)=\int N_{a}^{i}{\cal C}_{i}^{a}$, $G(\alpha)=\int \alpha^{i} {\cal G}_{i}$.
We see that they form a f\/irst class system, thus they reduce the phase space by $2\times 3N=6N$ dimensions, leaving a zero-dimensional space of solutions. Hamiltonian systems with linearly dependent constraints are called reducible~\cite{BlauTFT}.

The symmetry group  also includes the action of dif\/feomorphisms, which manifestly leaves~\eqref{SBF} invariant. Dif\/feomorphisms are expressed as special combinations of gauge and shift transformations \cite{BuffenoirPleb, FreidelLouapre}. Recalling that the Lie derivative of a 1-form verif\/ies ${\cal L}_\xi\om=i_\xi d\om+d(i_\xi\om)$, where $i_\xi$ denotes the inner product with the vector $\xi$, it is easily checked that
\begin{gather*}
\d^D_\xi B = d^G_{i_\xi\om} B + \d^S_{i_\xi B} B + i_\xi(d_\om B),
\qquad \d^D_\xi \om = d^G_{i_\xi\om} \om + \d^S_{i_\xi B} \om + i_\xi F.
\end{gather*}
Accordingly, the canonical generators of hypersurface deformations can be expressed as linear combinations of~\eqref{BFcon}.

In spite of the lack of local dynamics, BF theory has many interesting applications. The subject of this review is the classical relation to general relativity, but the action has been used also in connection with Yang--Mills theory (e.g.~\cite{CattaneoBFYM}). Concerning the quantum theory, we refer the reader to~\cite{BlauTFT} for the continuum path integral quantization and its relation to the measure of f\/lat connections and Reidemeister torsion, and to~\cite{BaezIntro,Barrett:2008wh,Bonzom:2011br,PerezLR, Zapata:1996ij} for the discrete path integral and its relation to spin foam models for quantum gravity.

\section{Self-dual Plebanski action}

A way to relate BF theory to general relativity is to work with the local gauge group SU(2), seen as a chiral subgroup of the Lorentz group. The fundamental f\/ields are $(B^i,\om^i)$, with $i=1,2,3$ indices in the adjoint representation.
Then one can write a densitized symmetric tensor as~\cite{Capo2, Urbantke}
\begin{gather}\label{gU}
\sqrt{|g|} \, g_{\mu\nu} = \f1{12} \eps_{ijk}  \eps^{\a\b\g\d}
B^{i}_{\mu\a}B^{j}_{\b\g}B^{k}_{\d\nu}.
\end{gather}
A theorem by Urbantke states that if the bilinear density $m^{ij}\equiv B^i \w B^j$ is invertible as a matrix in the $ij$ indices, then also the Urbantke metric \eqref{gU} is invertible, and furthermore $B$ is self-dual (for $\det m>0$, else antiself-dual) with respect to it, that is $B^i_{\mu\nu}=\pm 1/(2\sqrt{|g|})\eps_{\mu\nu}{}^{\rho\sigma} B^i_{\rho\sigma}$.
See e.g.~\cite{FreidelWithout} for a recent proof.
Therefore, provided the $B$ f\/ield is not degenerate in the above sense, we have a natural way to introduce an invertible metric as a composite object.
For $B$ real, compatible with the case of Euclidean signature, one automatically obtains a real metric with positive determinant\footnote{Notice that this includes the case of Kleinian signature $(++--)$. The latter and the Euclidean case are distinguished by the signature of~$m^{ij}$, which is respectively Lorentzian and Euclidean \cite{FreidelWithout}.}.
For $B$ complex, as required by the chiral splitting of the Lorentz group with Lorentzian signature, one obtains a complex metric with negative determinant, and additional reality conditions need to be given \cite{Capo2}.

To deal with this formalism, it is useful to introduce a tetrad $e^I$ for the metric \eqref{gU},
\begin{gather}\label{deftetrad}
g_{\mu\nu} = e^I_\mu e^J_\nu \eta_{IJ},
\end{gather}
and def\/ine the Plebanski 2-forms $\Sigma^i_{\eps}\equiv P_{\eps}{}^i_{IJ} e^I \w e^J= e^{0}\wedge e^{i} + \sqrt{\sigma}\eps/2 \eps^{i}{}_{jk} e^{j}\wedge e^{k}$, where $\sigma=\pm$ is the spacetime signature,
$\eps=\pm$ and $P_{\scr(\pm)}{}^a_{IJ}$
are the projectors on the left- and right-handed $\mathfrak{su}(2)$ subalgebras, according to the isomorphism $\mathfrak{so}(3,\sigma)\cong\mathfrak{su}(2)\oplus\mathfrak{su}(2)$, and we have f\/ixed the time gauge in the internal indices. These forms have the property that the left-handed part (resp.\ right-handed) is simultaneously self-dual (resp.\ antiself-dual) in the spacetime indices with respect to~$e$.
The set $\Sigma^a_{\eps}$ provides an orthogonal basis for the 6-dimensional space of 2-forms. Combining this fact with Urbantke's theorem, we conclude that a generic, albeit non-degenerate, $B$ f\/ield can be written as a linear combination of Plebanski's 2-forms. This can be conveniently parametrized as follows,
\begin{gather}\label{BbS}
B^i_{\scr{(\pm)}} = \eta b^i_a \Sigma^a_{\scr{(\pm)}}, \qquad \det b^i_a = 1, \qquad \eta=\pm,
\end{gather}
where the unimodularity of $b^i_a$ can always be we assumed, as we did here, thanks to the conformal invariance of the notion of self-duality.
The parametrization \eqref{BbS} helps to appreciate the relation between SU(2) BF theory and general relativity:
it shows how the $B$ f\/ield can equivalently  be parametrized in terms of a metric $g_{\mu\nu}$ together with the eight components of the unimodular ``internal triad'' $b^i_a$, and a sign.
From the ``triad'' $b^i_a$, one can def\/ine the unimodular ``internal metric'' $q_{ab} := b^i_a b^j_b \d_{ij}$, given by an SL(3)-rotated form of the identity.
Comparing \eqref{gU} with \eqref{deftetrad}, we see that the $B$ f\/ield plays the role of a chiral ``cubic root'' of the metric, with the SL(3) group formed by the $b^i_a$ replacing the internal Lorentz group in \eqref{deftetrad} \cite{thooftBF}\footnote{This picture can be further extended to the whole GL(3) group \cite{Bengtsson:1995rz}.}.

The relation with general relativity can now be easily established:
It suf\/f\/ices to impose the condition that $b^i_a\in$ SO(3), or equivalently that
$q_{ab}=\d_{ab}$,
and the BF action \eqref{SBF} immediately reduces to ($\eta$ times) the Einstein--Cartan action in self-dual variables, with cosmological constant $\Lambda=-3\eta\lambda$.
The required condition on $b$ can equivalently be written as
\begin{gather}\label{metrC}
B^i\w B^j = \f13 \d^{ij} B_k\w B^k,
\end{gather}
a form known as Plebanski or metricity constraints.
Using a traceless symmetric scalar $\varphi_{ij}$ as a~Lagrange multiplier, the constraints can be included in the action,
\begin{gather}\label{PlebAction}
S(B,\om) = \int B_i \w F^i(\om) -\f{\lambda}2   B_i\w B^i + \f12\varphi_{ij}   B^i\w B^j.
\end{gather}
This is Plebanski's self-dual formulation of general relativity \cite{Capo2,MikeLeft}, to which it is equivalent in the sector of non-degenerate $B$ f\/ields\footnote{This is a condition that does not change the dimensionality of the conf\/iguration space, by all means similar to requiring the invertibility of the metric in Einstein--Cartan.}.
The role of the constraints is to freeze the scalar degrees of freedom $b^i_a$, and leave only the Urbantke metric as a dynamical f\/ield.
In the case of Lorentzian signature, one takes  complex $B$ f\/ields, and adds reality conditions on the metric~\cite{Capo2}.

It is also instructive to see the equivalence at the level of the f\/ield equations.
To that end, let us f\/ix the solution $B=\eta\Sigma_{(\scr +)}(e)$ to the constraints (the equivalence in the other sectors is analogous to what follows). The compatibility condition
$d_{\om} B=0$ is then solved by the (left-handed part of the) spin connection $\om(e)$, and its curvature yields the left-handed part of the Riemann tensor,
$F^i_{\mu\nu}(\om(e))=P^i_{(\scr +)}{}_{\rho\sigma}R^{\rho\sigma}{}_{\mu\nu}(e)$. As such, it can be decomposed into its trace-free Weyl part, and its Ricci part, by projecting also the second pair of indices,
$F^i_{\mu\nu}(\om(e))
=f^i{}_a P^a_{(\scr +)}{}_{\mu\nu} + \bar f^i{}_a P^a_{(\scr -)}{}_{\mu\nu}$. The symmetric matrix $f^i{}_a$ contains the f\/ive Weyl components plus the trace of Ricci, and the generic matrix $\bar f^i{}_a$ contains the nine components of the trace-free Ricci tensor. We now look at the remaining f\/ield equations,
\begin{gather*}
F^i=\lambda B^i-\varphi^i_j B^j.
\end{gather*}
Using the above decomposition of $F$ and the self-duality of~$B$, these equations are equivalent to $\bar f^i{}_a=0$, ${\rm tr} \,f^i{}_a=3\eta\lambda$, with the remaining components of $f^i{}_a$ free. These are precisely the Einstein's equations, namely
$R_{\mu\nu}-\f14g_{\mu\nu}R=0$, $R=-4\Lambda$, Weyl free, with $\Lambda=-3\eta\lambda$.

The use of self-dual variables has been a fertile ground for a number of applications in general relativity, especially in the study of exact solutions and integrability, see e.g. \cite{Dunajski:2010zz}.

{\bf Canonical analysis.}
The canonical analysis of the theory \cite{Capo2} is similar to the BF case treated earlier. The canonical variables are the same pair
$(\om_a^i, \Pi^a_i:=\f12\eps^{abc}B_{ibc})$, the system is fully constrained, and the constraints now read
\begin{gather*}
{\cal G}_i=D_a \Pi^a_i = 0,
\qquad {\cal C}^a_{i}= \f12 \eps^{abc} F_{ibc}(\om) -\lambda\Pi^a_i +\varphi_{ij}\Pi^{aj}=0,
\\ {\cal S}^{ij} = \Pi^{a(i} B^{j)}_{0a} - \f13 \d^{ij}\Pi^{a}_k B^{k}_{0a}=0.
\end{gather*}
The main novelty is the fact that together with the Hamiltonian, the metricity constraints ${\cal S}$ form a second class system, that is, their Poisson brackets do not vanish on the constraint surface.
In this situation, one typically proceeds solving explicitly the part of the constraints which is  second class, and derive a reduced action in terms of only f\/irst class constraints.
Assuming that the density-weight 1 $\Pi^a_i$ has an inverse, $\Pi^a_i \Pi^i_b = \d^a_b$, we can solve them  by f\/ixing the value of the Lagrange multiplier~$B_{0a}^{i}$
\[
B^i_{0a} = \sqrt{\det \Pi}\big(\d^{ij} N + \eps^{ijk} N_{k} \big) \Pi_{a j} .
\]
Plugging the result back into the action, one f\/inds
\[
S = \int \Pi^a_i \dot{\om}^i_a + \om_0^i {\cal G}_i + N {\cal H}+ N^a {\cal H}_a,
\]
where $\cal G$ is the same Gauss constraint as before, ${\cal H}_a = \Pi^b_i F^i_{ab}$ is the vector constraint, and
\[
{\cal H} = \f1{\sqrt{\det \Pi}} \eps_{ijk} F^i_{ab} \Pi^{aj} \Pi^{bk} - \lambda   \sqrt{\det \Pi}
\]
is the scalar Hamiltonian constraint.
The result is the Ashtekar Hamiltonian~\cite{Ashtekar:1987gu} in its original complex version, with the self-dual part of the Lorentz connection as the fundamental variable.

Interestingly, the structure of the constraint algebra, and with it the number of physical degrees of freedom, is unchanged if one replaces the constant~$\lambda$ by a functional $\lambda(\varphi^{ij})$ where~$\varphi^{ij}$ is the symmetric, trace-free spacetime scalar built from
$ \eps^i{}_{kl} F^j_{ab} \Pi^{ak} \Pi^{bl}$. This was noticed in~\cite{Bengtsson:1990qg, Peldan},
and has recently been pointed out again in~\cite{Krasnov2, Krasnov1}. We will come back to it below.

{\bf Quantization.}
The canonical quantization of this theory is the original program of loop quantum gravity in complex variables~\cite{Ashtekar:1987gu,ThiemannBook}, where
one looks for an Hilbert space of functionals of the Ashtekar self-dual connection.
Notoriously, the main dif\/f\/iculty with this approach lies in the reality constraints. The reality conditions~\cite{Capo2} are quadratic in the f\/ields, and pose no particular problem at the classical level. However, they have so far provided a stumbling block for the quantum theory, and the program has then switched towards the use of a real connection, which is related to the non-chiral Plebanski action to be discussed below.
For the Euclidean case, where reality conditions are not needed, a simplicial path integral was introduced in \cite{MikeLeft}.

{\bf Second order BF action.}
One remarkable fact about SU$(2)$ BF theory is that it is always possible to explicitly  and uniquely solve the compatibility condition $d_{\om}B=0$, as long as
$B$ is assumed to be non degenerate, even if the simplicity constraint is not satisf\/ied. There exist  a~unique  compatible connection $\om(B)$ \cite{Bengtsson:1995rz, DeserTeit,Halpern,KrasnovEff},
given by
\begin{gather}\label{omB}
\om^i_\m(B) = \f1{4e} \eps^{\r\s\lambda\tau} B^{i}_{\lambda\tau} B_{j \r\m} \na^\n B^j_{\n\s}.
\end{gather}
Here the indices are raised and lowered with the Urbantke metric, $e$ is the determinant of its tetrad and $\na$ its covariant derivative.
Using this result, one can give a second order formulation of the BF action~\eqref{SBF}. This formulation becomes particularly interesting if one uses the parametrization~\eqref{BbS} for~$B$
in terms of the Urbantke metric and the scalar f\/ields~$b_{a}^{i}$ or their metric~$q_{ab}$. This was f\/irst proposed in~\cite{FreidelWithout} and  in that case, the curvature of~\eqref{omB} includes the Riemann tensor of the metric.
Therefore, one can obtain a second order formulation of SU(2) BF theory as an action for a metric and scalar f\/ields, without any constraint needed.
The result can be conveniently written using the internal unimodular metric $q_{ab}$ as follows~\cite{FreidelWithout},
\begin{gather}\label{SF}
S(e^I_\mu, q_{ab}) = \f{\eta\eps}4 \int e R_\eps^{ab}(e) (\hat q \d_{ab} - \hat q_{ab})
+ \f12 e \nabla_\eps^\mu q_{ab} C_{\eps \ \mu\nu}^{abcd} \nabla_\eps^\nu q_{cd}
-\eps \lambda \int e q.
\end{gather}
Here  $\hat q_{ab}$ is the inverse of $q_{ab}$, with trace $\hat q$ (internal indices are still raised and lowered with the identity metric),
\begin{gather}\label{Riem}
R^{ab}_\eps(e)=\f12 \Sigma^a_\eps{}_{\mu\nu}(e)\Sigma^b_\eps{}_{\rho\sigma}(e)R_{\mu\nu\rho\sigma}(e)
\end{gather}
is the self- or antiself-dual (resp.\ for $\eps=\pm$) part of the Riemann tensor, $\nabla_\eps{}_\mu$ is the covariant derivative with respect to the spin connection $\g_\eps^a{}_b(e)=\eps^a{}_{bc} P^c_\eps{}_{IJ}\om^{IJ}(e)$, and f\/inally
the kernel of the kinetic term is given by
\begin{gather*}
C^{abcd}_{\eps \ \mu\nu}(e^I_\mu, q_{ab}) \equiv \left(\d^{ad}\d^{bc} - \f12 \d^{ab}\d^{cd} \right) g_{\mu\nu}
+ \left(\d^{bc} \eps^{ad}{}_g - \hat q^{bc} \eps^{adf} q_{fg} \right) \Sigma_\eps^g{}_{\mu\nu}(e).
\end{gather*}
The form \eqref{SF} shows that BF theory, which is topological, can be formulated in terms of the Riemann tensor~\eqref{Riem} of a space-time metric,
coupled to scalar f\/ields, that form the component of a internal SL(3) metric, which posses their own dynamics. The presence of these extra terms is crucial to have the shift symmetry~\eqref{st}, which is what eliminates any local degrees of freedom present a priori in the metric and the $b$ scalars. The metricity constraint~\eqref{metrC} then plays a double role: on the one hand, it freezes the scalars. On the other hand, it is responsible for breaking the shift symmetry, leaving the local gauge and dif\/feomorphisms.
Imposing the metricity constraint $q_{ab}=\d_{ab}$, it is easily verif\/ied that~\eqref{SF} reduces to
the Einstein--Hilbert Lagrangian $\eta\eps/2 e [R(e)-6\eta\lambda]$.

This construction also gives a geometric interpretation to the scalars~$b$. In fact,
one can use the triad $b$ to introduce a new connection acting on the internal indices, as
\[
A^a{}_b \equiv
\hat b^a_j d_\om b^j_b.
\]
It is then easy to show that the new connection satisf\/ies
\[
d_A\Sigma^a_\eps(e)=0, \qquad d_A q_{ab}=0.
\]
That is, $A$ is an SL(3) connection, torsion-free, but non-metric with respect to the local structure group~SU(2).
In this sense, the~$b$ describe the non-metricity of the torsion-free connection~$A$.

\looseness=-1
{\bf Modif\/ied theories of gravity.}
The discussion above pointed out the importance of brea\-king the shift symmetry in order to get propagation of local degrees of freedom.
A generic way of doing so is to add a potential term for $\phi$ to~\eqref{PlebAction}, or equivalently, a potential for $B$ to the BF action,
\begin{gather}\label{modPleb}
S(B,\om) = \int B_i \w F^i(\om) + \left(V(B)-\f{\lambda}2\right)   B_i\w B^i.
\end{gather}
The potential $V(B)$ breaks the shift symmetry, leaving dif\/feomorphisms and local gauge transformations as symmetries of the action.
Notice that it has to be a functional of (scalar contractions of) $B^i\w B^j$, so ef\/fectively $V(B) = V(q, q_{ab}^2)$.
This class of actions has been introduced by Krasnov~\cite{Krasnov1}, and is related to the earlier formulations~\cite{Bengtsson:1990qg,Capo3,Peldan}.
In the light of~\eqref{SF}, we see that the result is an action for a metric and additional scalar f\/ield, with dynamics dif\/ferent from general relativity and driven by the precise form of~$V(B)$.

An appealing aspect of \eqref{modPleb} is that both BF and Plebanski's actions are special cases thereof, obtained respectively for a vanishing potential, and for a singular potential imposing \eqref{metrC} as a~constraint. As emphasized e.g.\ in~\cite{KrasnovEff}, this fact provides a well-def\/ined scheme where BF theory could arise as the ultra-violet f\/ixed point of low-energy general relativity, a possibility entertained in the literature (e.g.~\cite{Rivasseau:2011xg}).
But what is really remarkable, is that for any potential, the theory still describes only two degrees of freedom. This can be shown through a canonical ana\-ly\-sis~\mbox{\cite{Krasnov2, Krasnov1}}, which gives exactly what was anticipated in a previous subsection.
The result makes~\eqref{modPleb} unlike other modif\/ications of general relativity: In actions with a metric tensor as the unique f\/ield, modif\/ications require the introduction of higher order dif\/feomorphism invariants,
which in turn carry higher order derivatives, leading in general to extra degrees of freedom~\mbox{\cite{Deruelle, Stelle}}.
What allows the absence of extra degrees of freedom in this context is the  presence of an \emph{infinite} number of higher derivatives. Indeed, as shown f\/irst in~\cite{FreidelWithout} and furtherly developed in~\cite{KrasnovEff}, the~$b$ scalars can be integrated out, order by order in perturbation theory, thus obtaining a theory of the metric alone and with only two degrees of freedom, but at the price of generating an inf\/inite number of higher derivatives terms.
Some possible applications of this class of actions, both as modif\/ied classical theories of gravity, as grand unif\/ied theories, and as quantum mechanical ef\/fective theories, are discussed in the literature~\cite{IoAki,KrasnovEff,Krasnov:2010tt, Krasnov:2008sb}.

\section{Non-chiral Plebanski action}

The Plebanski action can be formulated also using the full Lorentz group ${\rm SO}(3,\sigma)$ \cite{Capo2,DePietri,MikeLR}.
In this non-chiral, or covariant, version the fundamental f\/ields are $(B^{IJ},\om^{IJ})$, $I=0,\ldots, 3$.
The key expression \eqref{gU} for the Urbantke metric can be straightforwardly generalized to this case, but now \emph{two} possible metrics exist. This is simply a consequence of the fact that the tensor product of three adjoint representations of the algebra admits two singlets.
A basis in this two dimensional vector space is provided by the tensors
$\eta_{N[I} \eta_{J]MKL}$ and $\eta_{N[I} \eps_{J]MKL}$, where $\eps_{IJKL}$
is the completely antisymmetric tensor and we def\/ined the identity $\eta_{IJKL} = \f12 (\eta_{IK} \eta_{JL}-\eta_{IL} \eta_{JK})$. Accordingly, we have a right-handed Urbantke metric $g^{(+)}_{\mu\nu}$ and a left-handed $g^{(-)}_{\mu\nu}$,
\begin{gather}\label{gpm}
\sqrt{g^{(\pm)}}   g^{(\pm)}_{\mu\nu} =
\f1{12} \eta_{IN}\left(\eta_{JMKL} \pm \f{\sqrt{\sigma}}2 \eps_{JMKL}\right) \eps^{\a\b\g\d} B^{IJ}_{\mu\a}B^{KL}_{\b\g}B^{MN}_{\d\nu},
\end{gather}
and $\sigma=\pm 1$ is the spacetime signature.

$B^{IJ}$ can be immediately parametrized in terms of these two Urbantke metrics, using the left/right decomposition and applying twice~\eqref{BbS}.
As the right- and left-handed parts are independent, the two sectors have independent triads and tetrads, say
$b^i_a$, $\bar b^i_a$ and $e^I_\mu$, $\bar e^I_\mu$. Correspondingly, we have
\begin{gather}\label{ParamGen}
B^{IJ} =  P^{IJ}_{\scr (+)}{}_i   b^i{}_a \Sigma^a_{\scr(+)}(e)
+ \eta P^{IJ}_{\scr (-)}{}_i   \bar b^i{}_a \Sigma^a_{\scr(-)}(\bar e),
\end{gather}
where the Plebanski 2-forms encode the two metrics
$g_{\mu\nu}^{(+)}=e^I_\mu e^J_\nu \d_{IJ}$ and $g_{\mu\nu}^{(+)}= \eta \bar g_{\mu\nu} = \eta \bar e^I_\mu \bar  e^J_\nu \d_{IJ}$, and we dropped an irrelevant overall sign.

In the previous SU(2) case, the reduction of the 18 components of $B$ to a metric up to an SU(2) rotation required 5 constraints, see~\eqref{metrC}. We now have 36 initial components, and 6 internal gauge freedoms. We thus need 20 constraints: 10 to freeze the $b$ and $\bar b$ scalars, and 10 to identify $g$ and $\bar g$ into a unique metric.
This can be immediately achieved with a covariant version of~\eqref{PlebAction}~\cite{Capo2,DePietri,MikeLR,MikeNewC}
\begin{gather}\label{PlebActionC}
S(B,\om) = \int B_{IJ} \w F^{IJ}
+ \f\Lambda4  \eps^{IJ}_{KL}   B_{IJ}\w B^{KL} + \f12 \phi_{IJKL}   B^{IJ}\w B^{KL}.
\end{gather}
The Lagrange multiplier $\phi_{IJKL}$ is by def\/inition symmetric under the exchange of the f\/irst and second pair of indices. Since it is also antisymmetric within each pair, this leaves 21 independent components. This is one too many, thus one needs to remove one component if we want twenty constraints. The standard choice in the literature is to take the additional condition $\eps^{IJKL}\phi_{IJKL}=0$.
The multiplier then decomposes into irreducible representations as
$
\phi \in {\bf (2,0)}\oplus{\bf (0,2)}\oplus{\bf (1,1)}\oplus{\bf (0,0)},
$
giving rise to corresponding constraints, known as simplicity constraints.
For our purposes, it is suf\/f\/icient to give an overview of their role, without looking at their explicit form.
The f\/ive-dimensional self-dual representation ${\bf (2,0)}$ is again the constraint~\eqref{metrC} for the purely self-dual part of $B^{IJ}$, thus it freezes the $b$ scalars in~\eqref{ParamGen}. Similarly, the ${\bf (0,2)}$ freezes the~$\bar b$.
The ten-dimensional constraints in the representations ${\bf (1,1)}\oplus{\bf (0,0)}$ impose the coincidence of the tetrads, $e_\mu^I\equiv \bar e^I_\mu$, by requiring the orthogonality of~$\Sigma_{\scr (+)}(e)$ and $\Sigma_{\scr (-)}(\bar e)$.
See~\cite{bimetric} for details.
Finally, the sign~$\eta$ remains free. Hence, the Urbantke metrics coincide up to a~sign, $g_{\mu\nu}^{(+)}=\eta g_{\mu\nu}^{(-)}$. Plugging this result in \eqref{ParamGen}, the two sectors $\eta=\pm1$ give respectively
\begin{gather}\label{Bsols}
B^{IJ}=e^I\w e^J, \qquad B^{IJ}=\f12\eps^{IJ}_{KL} e^K\w e^L.
\end{gather}

The theories described by these solutions can be examined simply by looking at the reduced actions for $e$ and $\om$, which read respectively
\begin{subequations}\label{SEC}
\begin{gather}
S_{\rm top}\big(e^I_\mu,\om^{IJ}_\mu\big) = \int \tr [e \w e \w F(\om)]+6\Lambda e, \\
S_{\rm EC}\big(e^I_\mu,\om^{IJ}_\mu\big) = \int \tr [\star e \w e\w F(\om)]+6\Lambda e,
\end{gather}
\end{subequations}
where $\star=1/2 \eps^{IJ}_{KL}$.
The f\/irst action describes a topological theory, with no local degrees of freedom, see e.g.~\cite{Liu}.
The second is the Einstein--Cartan action for general relativity. This is how the non-chiral Plebanski action describes gravity. See also~\cite{SmolinSpeziale} for the equivalence at the level of f\/ield equations.

With respect to the self-dual action, the non-chiral formulation has the advantage that no additional reality conditions are needed in the Lorentzian case.
In fact, although the individual Urbantke metrics~\eqref{gpm} are complex when $B$ is real (an $i$ appears in front of the epsilon tensor), the f\/inal tetrad~\eqref{Bsols} is automatically real once the simplicity constraints are imposed.
The other important aspect of this formulation is to be related to loop quantum gravity in real variables, as we now review.

{\bf Including the Immirzi parameter.}
The Lagrangians appearing in \eqref{SEC} can be considered in a unique action principle,
\begin{gather}\label{SH}
S = \int {\rm tr} [P_\g \, e \w e \w F(\om)] + 6\Lambda e, \qquad P_\g = \f1\g + \star,
\end{gather}
where the coupling constant $\g$ can be identif\/ied with the Immirzi parameter of loop quantum gravity, up to a topological term \cite{Date}.
The Plebanski action \eqref{PlebActionC} can be easily adapted to reduce to \eqref{SH}. The simplest way to do so is to include the second invariant in the kinetic term of \eqref{PlebActionC}, that is make the replacement
\[
{\tr} [B\w F] \ \to \ \tr [P_\g   B\w F].
\]
Now both solutions \eqref{Bsols} give an action like \eqref{SH}.
Alternatively, one can choose the missing component of the Lagrange multiplier to be a linear combination of the two scalars $\d^{IJKL}\phi_{IJKL}$ and $\eps^{IJKL}\phi_{IJKL}$. Then \eqref{Bsols} is replaced by the unique solution $B=P_\g \, e\w e$, and again an action like \eqref{SH} is recovered. See \cite{CapoBFImmi,SmolinSpeziale} for discussions of these alternative cases. For more on the non-chiral Plebanski action, see e.g.~\cite{Baratin:2011hp,Montesinos06Altern,MontesinosVelazquez,Perez:2002vg,MikeNewC,Wieland1}.

{\bf Canonical analysis.}
The canonical analysis of this theory has again second class constraints, but the analysis is considerably more involved than for the self-dual case.
We do not present the details here, rather refer the reader to \cite{AlexandrovKrasnov,BuffenoirPleb,Wieland1}. Again, the solution of the second class constraints, with or without time gauge, is given by an SU(2) connection, the Ashtekar--Barbero connection \cite{Barbero,Immirzi96Real}. See \cite{AlexandrovLivine,Barros,Cianfrani,GeillerNewLook, Holst}. This connection depends on the Immirzi parameter, and should be viewed as an auxiliary f\/ield, not straightforwardly related to the initial Lorentz group.
The resulting theory has the phase space of a non-abelian gauge theory, and marks the starting point of loop quantum gravity \cite{AshtekarReport,IoLectures, CarloBook,ThiemannBook}\footnote{An alternative solution has been proposed in \cite{AlexandrovChoice,AlexandrovBuffenoir}. See \cite{AlexandrovSigma} for a recent overview.}.
It is interesting to add that canonically, the simplicity constraints can also be given a linear version. This linear formulation is at the root of the new spin foam models \cite{EPRL, EPR,FK,LS2}.

Recently in \cite{Bodendorfer1,BodendorferSimpl}, it has also been proposed to approach the second class constraints via a~dif\/ferent procedure, known as ``gauge unf\/ixing'' \cite{Anishetty,HenneauxBook, Mitra}. That is, one views the second class constraints as the gauge-f\/ixing part of a larger system with only f\/irst class constraints. This procedure typically involves a more complicated Hamiltonian, however the results of \cite{Bodendorfer1,BodendorferSimpl} indicate that interesting new insights can be achieved this way, including a generalization of the loop quantum gravity techniques to higher dimensions and supergravity.

{\bf Quantization.}
The canonical quantization of the action \eqref{PlebActionC}, in a non-perturbative and background-independent way, is the program of loop quantum gravity in real variables. The same action is also the typical starting point for spin foam models \cite{PerezLR}.
The perturbative quantization of \eqref{SH} has been studied in \cite{IoDario,IoDarioProc}, bringing to light the renormalization of the Immirzi parameter. Possible roles of the Immirzi parameter have been considered in the coupling to fermions \cite{AlexandrovFermions, FreidelMinic,Mercuri,PerezRovelli}, and in cosmology~\cite{MagueijoComplexImmirzi}.

{\bf Relaxed constraints and bi-metric theories of gravity.}
The above discussion highlights the fact that the non-chiral Plebanski action is naturally a theory of two metrics. It is only the presence of the simplicity constraints that forces the two metrics to coincide.
One can consider a modif\/ication of the theory along the lines of \eqref{modPleb}, where the constraints are replaced by a~potential for $B$,
\begin{gather}\label{modPlebC}
S(B,\om) = \int B_{IJ} \w F^{IJ}
+ \left( V(B) + \f\Lambda4\right) \eps^{IJ}_{KL}   B_{IJ}\w B^{KL}.
\end{gather}
In the absence of constraints, the two Urbantke metrics are independent and dynamical. In this case, extra degrees of freedom can be expected. This was indeed shown in \cite{AlexandrovKrasnov}, where a~canonical analysis showed that for generic potentials, one has eight degrees of freedom. This was further clarif\/ied in \cite{bimetric}, where using \eqref{ParamGen} and \eqref{SF}, it was shown that the modif\/ied Plebanski theory is equivalent to a bi-metric theory of gravity, plus the additional $b$ and $\bar b$ scalars. The interpretation of the eight physical degrees of freedom turned out to be a massless graviton, a~massive spin-2 particle, and a scalar\footnote{This identif\/ication is made via a perturbative expansion around the ``bi-f\/lat'' background,
$g_{\mu\nu}=\eta_{\mu\nu}+h_{\mu\nu}$,  $\bar g_{\mu\nu}=\eta_{\mu\nu}+\bar h_{\mu\nu}$.
Then one can consider the following linear combinations, $h^{\pm}_{\mu\nu}= h_{\mu\nu}\pm \bar h_{\mu\nu}$. Since $h^-_{\mu\nu}$
is invariant under dif\/feomorphisms, its masslessness is no longer protected by the symmetries. A generic potential term in~\eqref{modPlebC} will result in a non-zero mass. Such a f\/ield in general propagates 6 degrees of freedom, corresponding to a spin-2 and a spin-0 particle. See~\cite{bimetricPF, bimetric} for more details.}. If~\eqref{modPlebC} is extended to include the Immirzi parameter, the mass turns out to depend on it~\cite{bimetricPF}.

Interestingly, this is the same spectrum of pure bi-metric theories~\cite{Damour}. That is, one seems to be able to put all the relevant degrees of freedom in the metrics, without the scalars, likewise to the self-dual theory (see discussion in the previous section). On the other hand, the structure of a~generic potential, to be built out of scalar invariants of $Q^{IJKL}=B^{IJ}\w B^{KL}$, is now much richer.
In particular, classes of potentials with lesser than eight degrees of freedom can be identif\/ied. These includes for instance Pauli--Fierz mass terms, and simpler scalar-tensor theories. These aspects are investigated in~\cite{BekeProc, bimetricPF}, and we refer the reader to these references for further details, as well as discussions on the dif\/ferent reality conditions possible in the modif\/ied context.

Concerning applications, actions of this type can arise as ef\/fective descriptions of spin foam models of quantum gravity (e.g.~\cite{EteraHoloEucl}),
and have been also used in grand unif\/ied theories, see~\mbox{\cite{LisiLeeIo, SmolinPlebGUT}}, where it resonates with similar ideas reviewed in \cite{Percacci}.
More in general, bi-metric theories of gravity are interesting in the context of modif\/ied theories of gravity \cite{Clifton:2010hz}.

\section[MacDowell-Mansouri action]{MacDowell--Mansouri action}

A dif\/ferent mechanism is to introduce a tetrad through some components of the connection,
as it happens in the Chern--Simons formulation of 2+1 gravity~\cite{Witten3d}.
The specif\/ic choice of gauge group depends on the spacetime signature, and on the sign of the cosmological constant
one is interested in. In the case of a 4d Lorentzian spacetime, one takes as~$G$ the De Sitter group~SO(4,1)
for $\Lambda>0$, and the anti-de~Sitter group SO(3,2) for $\Lambda<0$. For Euclidean signature, the respective choices are~SO(5) and~SO(4,1).
To be specif\/ic, in the following we will consider Euclidean signature and positive~$\Lambda$, but the results are easily generalized. Seen as a~vector space, the Lie algebra splits as follows,
\[
{\mathfrak{so}(5)} = {\mathfrak{so}(4)} \oplus \R^{4},
\]
where the homogeneous space
$\R^{4}\simeq {\rm SO}(5) / {\rm SO}(4)$ is spanned by the generators of translations.
The splitting is orthogonal with respect to the Killing form, and ${\mathfrak{so}(4)}$-invariant.
It lets us introduce a tetrad as the vectorial part of the larger connection. To see it,
let $A^{\alpha\beta}$ be the initial~$\mathfrak g$ connection, $\alpha=0,\ldots, 4$, and let $I=0,\ldots, 3$.
We then introduce the components
\begin{gather}\label{iden}
A^{IJ} = \om^{IJ}, \qquad  A^{I4} = \f1\ell e^I.
\end{gather}
Here $\ell$ is a constant with dimensions of a length, necessary to give both the tetrad and the connection
canonical dimensions. It will turn out to be related to the cosmological constant via $\ell^2=3/\Lambda$.\footnote{An alternative formulation considers the possibility of interpreting $\ell$ as the Planck length \cite{Townsend1}, but we will not discuss this possibility here.}
That is, we view the SO(4) connection and the tetrad as two pieces of a unique~$\mathfrak g$ connection,
as one does for Chern--Simons gravity in 2+1.
A simple calculation gives
\begin{gather}\label{Faom}
F^{IJ}(A) = F^{IJ}(\om) -\f1{\ell^2} e^I \w e^J, \qquad F^{I4} = \f1\ell d_\om e^I.
\end{gather}

The equation $F^{\alpha\beta}(A)=0$ of BF theory alone thus implies de~Sitter spacetime as the unique non-degenerate metric solution. To f\/ind all the solutions of general relativity, one needs to break the symmetry down to SO(4). This can be done adding an extra term in the BF action, which introduces a direction in the internal space:
\begin{gather}\label{BFMCM}
S= \int B_{\a\b}\w F^{\a\b} -\f\lambda2 B_{\a\b}\w B^{\a\b}
- \f12 \eps_{\alpha\beta\gamma\delta\varepsilon} v^\varepsilon B^{\alpha\beta}\w B^{\gamma\delta}.
\end{gather}
The extra term breaks not only SO(5), but also the shift symmetry \eqref{st}, thus local degrees of freedom can be expected. Notice that with respect to the previous Plebanski formulation, we are adding here a potential-like term, and not a constraint.
Here $v^\alpha=(0,0,0,0,v)$, consistently with the identif\/ications~\eqref{iden}.

Since the action is quadratic in $B$, the f\/ield can be trivially eliminated using its f\/ield equations,
without any conditions on the remaining f\/ields $\om$ and $e$.
Substituting back into the action, one obtains
\begin{gather}\label{SMM}
S_{MM}(A) = \int \f{v}{2(v^2-\lambda^2)} \tr[\star F\w F] - \f{\lambda}{2(v^2-\lambda^2)} \tr[F \w F] + \f1\lambda F_{I4}\w F^{I4},
\end{gather}
where in the f\/irst two terms $F$ is the SO(4) curvature,
and the SO(4) Hodge start $\star$ is embedded in SO(5) via $\eps_{IJKL4}=\eps_{IJKL}$.
The f\/irst term is the MacDowell--Mansouri action for gravity, introduced in \cite{MacDowell} independently of the connection with BF theory. The remaining two are topological terms that do not af\/fect the f\/ield equations.
Hence, \eqref{BFMCM} provides a BF-like formulation of the MacDowell--Mansouri action.
The equivalence with general relativity is then established thanks to the identities \eqref{Faom}. After some simple algebra, one arrives at the form
\begin{gather}\label{Sgen}
S = \int \f1G \tr[P_\gamma  e\w e\w F(\om)] - \f\Lambda{6G} e + c_1 E(\om) + c_2 P(\om) + c_3 NY(e,\om),
\end{gather}
where the last three terms are the topological invariants Euler, Pontryagin and Nieh--Yan, and  the coef\/f\/icients are given explicitly by
\begin{gather*}
\g=\f\lambda{v}, \qquad G\Lambda =\f{3(v^2-\lambda^2)}{v},
\qquad c_1 = \f{v}{4(v^2-\lambda^2)} , \qquad c_2= -\f{\lambda}{2(v^2-\lambda^2)}, \qquad c_3=\f1\lambda,
\end{gather*}
where we identif\/ied  $\Lambda=3/\ell^2$.
Up to topological terms\footnote{It can also be observed that the six invariants in \eqref{Sgen} are the only ones that can be written without requiring invertibility of the tetrad, or auxiliary f\/ields.}, we obtained the Einstein--Cartan action for general relativity.
We remark that the action principle has no scale to begin with, only the dimensionless quantities $\lambda$ and $v$. A scale, $\ell$, is introduced when splitting the components of $A$ into connection and tetrad, and it sets the cosmological constant scale.

The procedure of splitting the components of the connection into a smaller connection and a~tetrad is reminiscent of the Chern--Simons formulation of three-dimensional gravity.
However, unlike Chern--Simons, the relevant action principle is \emph{not} invariant under the initial Lie algebra:
The symmetry is explicitly broken by the selection of a direction in the internal space.
Ef\/forts to try to make this dynamical have been discussed e.g.\ in~\cite{StelleWest, TseytlinPoincare} and \cite{Wilczek:1998ea}.
The typical dif\/f\/iculty is that the larger theory in which SO(5) is spontaneously broken possesses extra degrees of freedom,
and its stability and unitarity have not been studied.
An alternative viewpoint has been put forward in \cite{FreidelS}, where it was argued that the BF formulation~\eqref{BFMCM} exhibits a spontaneous symmetry breaking mechanism that comes from integrating in the path integral over the SO(5) gauge degrees of freedom.

Many aspects of this formulation of gravity have been studied in the recent literature. These include symmetry properties~\cite{Durka:2011yv}, a geometric interpretation of the de~Sitter local gauge group~\cite{Wise}, the study of generalized solutions~\cite{Randono:2009gy}, as well as grand unif\/ied theories~\cite{Lisi}.
Finally, the mechanism can be extended to the SL(5) group~\cite{Mielke:2011zz}.

{\bf Quantization.}
The action \eqref{SMM} looks promising for quantization, given its Yang--Mills form and dimensionless coupling constant. However, the situation is not so simple.
In fact, in the naive form~\eqref{SMM} in which the symmetry is explicitly broken, the action is de facto equivalent to~\eqref{Sgen} and quantization stumbles upon the usual dif\/f\/iculties.
On the other hand, it was suggested in~\cite{FreidelS} to take~\eqref{BFMCM} as the starting point, and consider a quantum path integral def\/ined as a~``topological'' expansion around the BF kinetic term. This idea is quite intriguing, but it is challenged by dif\/f\/iculties with the gauge-f\/ixing procedure: the zeroth order of the expansion has a dif\/ferent symmetry than the successive orders.
See also~\cite{IoTopExp} and~\cite{CattaneoBFYM} on this.

\section{Outlook}

One of the key dif\/f\/iculties with general relativity is the high non-linearity of its f\/ield equations. This complexity is  enhanced further in the Einstein--Hilbert action principle, which is non-polynomial in the fundamental f\/ield, the metric. To obtain a polynomial action, one has to expand the metric around a f\/ixed background. Then the perturbations can be quantized, but the theory is not renormalizable. An important line of research in quantum gravity imputes this failu\-re to the background-dependent, perturbative methods, and seeks a background-independent formulation. When seeking for alternative approaches, the use of dif\/ferent fundamental variables with simpler actions is a useful guiding principle. In this respect, the relation of general relativity with BF theory appears very promising.
The work appeared so far in the literature has unraveled the deepest level of such a classical relation, and introduced new tools and ideas to push forward the investigation of gravity in these variables.
These results can be of benef\/it to approaches such as loop quantum gravity and spin foam models.

\pdfbookmark[1]{References}{ref}
\LastPageEnding

\end{document}